\documentclass [12pt]{iopart}
\usepackage{graphicx}

\begin {document}

\title [PCS on Fe-122]{Point-contact spectroscopic studies on normal and superconducting AFe$_2$As$_2$-type iron-pnictide single crystals}
\author {Xin Lu$^1$, W. K. Park$^1$, H. Q. Yuan$^2$, G. F. Chen$^3$, G. L. Luo$^3$, N. L. Wang$^3$, A. S. Sefat$^4$, M. A. McGuire$^4$, R. Jin$^4$, B. C. Sales$^4$, D. Mandrus$^4$, J. Gillett$^5$, Suchitra E. Sebastian$^5$ and L. H. Greene$^1$  }
\address {$^1$ Department of Physics and Fredrick Seitz Materials Research Lab., University of Illinois at Urbana-Champaign, IL 61801}
\address {$^2$ Department of Physics, Zhejiang University, Hangzhou 310027, China}
\address {$^3$ Institute of Physics, Chinese Academy of Science, Beijing 100190, China}
\address {$^4$ Materials Science $\&$ Technology Division, Oak Ridge National Lab., Oak Ridge, TN 37831}
\address {$^5$ Cavendish Laboratory, Cambridge University, JJ Thomson Avenue, Cambridge CB3OHE, U.K.}
\ead {xinlu@illinois.edu}

\begin {abstract}
 Point-contact Andreev reflection spectroscopy (PCARS) is applied to
 investigate the gap structure in iron pnictide
 single crystal superconductors of the AFe$_2$As$_2$ (A=Ba, Sr) family (\textquotedblleft
 Fe-122\textquotedblright). The observed point-contact junction conductance
 curves, G(V), can be divided into two categories: one where Andreev
 reflection is present for both (Ba$_{0.6}$K$_{0.4})$Fe$_2$As$_2$ and
 Ba(Fe$_{0.9}$Co$_{0.1}$)$_2$As$_2$, and the other with a V$^{2/3}$
 background conductance universally observed extending even up to
 100 meV for Sr$_{0.6}$Na$_{0.4}$Fe$_2$As$_2$ and
 Sr(Fe$_{0.9}$Co$_{0.1})_2$As$_2$. The latter is also observed in point-contact junctions on the
 nonsuperconducting parent compound BaFe$_2$As$_2$. Mesoscopic phase-separated coexistence of
 magnetic and superconducting orders is
 considered to explain distinct behaviors in the superconducting
 samples. For Ba$_{0.6}$K$_{0.4}$Fe$_2$As$_2$, double peaks due to Andreev reflection
 with strongly-sloping background are frequently observed for point-contacts on
freshly-cleaved c-axis surfaces. If normalized by a background
baseline and analyzed by the Blonder-Tinkham-Klapwijk model, the
data show a gap size $\sim$3.0-4.0 meV with $2\Delta_0/k_BT_c
\sim$ 2.0-2.6, consistent with the smaller gap size reported in
the LnFeAsO family (\textquotedblleft Fe-1111\textquotedblright).
For the Ba(Fe$_{0.9}$Co$_{0.1}$)$_2$As$_2$, G(V) curves typically
display a zero-bias conductance peak.
\end {abstract}
\maketitle

\section {Introduction}
The recently-discovered iron-based superconductors
\cite{HHosono-1} have emerged as a whole new family of high
temperature superconductors, which are complimentary to the study
of cuprates. They have attracted intensive study in the community
focusing on their physical properties and potential applications.
Reminiscent of cuprates, the antiferromagnetic (AFM) phase in the
parent compounds is suppressed with chemical doping or higher
pressure, and it is proposed that superconductivity in iron
pnictides may originate from AFM fluctuations \cite{IMazin-1}. It
is pointed out that there are also vital differences in that the
AFM ground state in the cuprates and Fe-pnictides are
Mott-insulator and metallic, respectively. Soon after the discovery of Fe-pnicitides, ARPES and quantum oscillation measurements have revealed disconnected hole-like bands around the $\Gamma (0,0)$ point and
electron-like bands around the $M(\pi, \pi)$ point on the FS
\cite{HDing-1, HDing-2, AKaminski-1,XJZhou-1, Sebastian-1}. The
order parameter (OP) symmetry in the cuprates has been determined
to be $d_{x^2-y^2}$, while it is still not known for the iron-pnictides. A
first step in unveiling the mechanism of the new family of
superconductors is to determine the pairing symmetry. An extended
s-wave gap structure with a sign reversal ($s\pm$) on different
FSs is proposed by I. Mazin \cite{IMazin-1} where the hole-like
and electron-like bands are fully-gapped and have a $\pi$ phase
shift in the superconducting state. A fully-gapped state has been
confirmed by different experimental techniques such as ARPES
\cite{HDing-1, HDing-2, AKaminski-1,XJZhou-1}, penetration depth
\cite{KHashimot-1}, $\mu$SR \cite{AAAczel-1, AJDrew-1}, Hc$_1$
\cite{HHWen-3} and specific heat measurements \cite{HHWen-1}.
However, a universal power law rather than exponential behavior is
observed in the London penetration depth measurement for
RFeAsO$_{0.9}$F$_{0.1}$ (R=Nd, La) \cite {RProzorov-1}, and
Ba(Fe$_{1-x}$Co$_x$)$_2$As$_2$ at various doping levels
\cite{RProzorov-2, RProzorov-3}. The Hebel-Slichter
coherent peak is absent in NMR measurements \cite{NJCurro-1,
KMatano-1} and it is argued that is a natural consequence
of the extended $s\pm$ model in the superconducting state
\cite{DParker-1}.

 Point-contact Andreev reflection
spectroscopy (PCARS) is a powerful tool to investigate the gap
structure and OP symmetry in superconductors. A good example is
the case of MgB$_2$, where two gaps, originating from $\sigma$
and $\pi$ bands, are clearly present in the
point-contact spectra, and also through directional PCARS measurements, the gap dependences on temperature and magnetic field are systematically revealed \cite{Gonnelli-2, Gonnelli-Mgb2,
Naidyuk-MgB2}. Park $\emph{et al.}$ successfully applied PCARS to
reveal the $d_{x^2-y^2}$ symmetry of the superconducting OP in
CeCoIn$_5$ \cite{wkpark-2}. Shortly after the discovery of the
superconducting LnFeAsO (Fe-1111) family, some PCARS measurements
have been carried out, but the results are not yet conclusive,
partly due to the polycrystalline nature of the samples. Chen $\emph{et al.}$ report a conventional BCS-like superconducting gap with $2\Delta_0/k_BT_c \sim 3.7$ for
SmFeAsO$_{0.85}$F$_{0.15}$ \cite{CLChien-1}, while multiple gaps
are claimed by other groups with different detailed structures  in
the initial stage \cite{HHWen-2, Gonnelli-1, PSamuely-1, LCohen-1,
Gonnelli-3}. Among those who claim multiple gaps, some merging
agreements are being reached where 2$\Delta_1/k_B T_c\sim$2-3 and
2$\Delta_2/k_B T_c\sim$7-9. In this paper, we apply PCARS to
different AFe$_2$As$_2$-type superconducting single crystals to
elucidate their gap structure and OP symmetry.

\section {Experiment}
AFe$_2$As$_2$-type (Fe-122) parent compound and various superconducting single
crystals are grown out of FeAs flux by a high
temperature solution method, such as parent compound BaFe$_2$As$_2$($T_N\sim$135 K), electron-doped superconducting A(Fe$_{0.9}$Co$_{0.1}$)$_2$As$_2$(A=Ba or
Sr, $T_c\sim$22 K) \cite{DMandrus-1}, and hole-doped superconducting (Ba$_{0.6}$K$_{0.4}$)Fe$_2$As$_2$
($T_c\sim$37 K) and (Sr$_{0.6}$Na$_{0.4}$)Fe$_2$As$_2$ ($T_c\sim$35
K)\cite{NLWang-1}. The crystals have natural c-axis facets and are
cleaved in the air to expose fresh and shiny surfaces before
mounting on the sample holder. A sharp Au tip prepared by
electrochemical etch \cite{WKPark-1} is engaged onto the sample surface with a
conventional differential micrometer mechanism \cite{XinLu-phd}.
The contact between the Au normal metal and Fe-122 crystals is
made after cooling to $\sim$2 K. The junction conductance
$dI/dV=G$, as a function of the bias voltage $V$, is recorded by
the standard ac lock-in technique, where the superconductor is
always biased positively. Contacts may be lost due to
vibration or instability in thermal contraction. Here we report data
 obtained after dozens of contacts have been made on each
crystal, and some with changing temperature up to the bulk T$_c$
and magnetic field up to 9 Tesla. 

\section {Results and Discussions}

Two different categories of G(V) curves are observed for
point-contacts on these Fe-122 superconducting crystals: (1) For (Ba,K)Fe$_2$As$_2$
and Ba(Fe,Co)$_2$As$_2$, Andreev reflection signals are observed but distinct from each other; and (2) For (Sr,Na)Fe$_2$As$_2$ and Sr(Fe,Co)$_2$As$_2$, in the absence of
Andreev reflection, a universal power law behavior is seen. The latter is also observed for the
point-contact on the non-superconducting parent compound
BaFe$_2$As$_2$. Results and discussions are presented here for
each case separately.

\subsection{(Ba,K)Fe$_2$As$_2$}
\begin{figure}[!ht]
 \includegraphics[angle=0,width=0.95\textwidth]{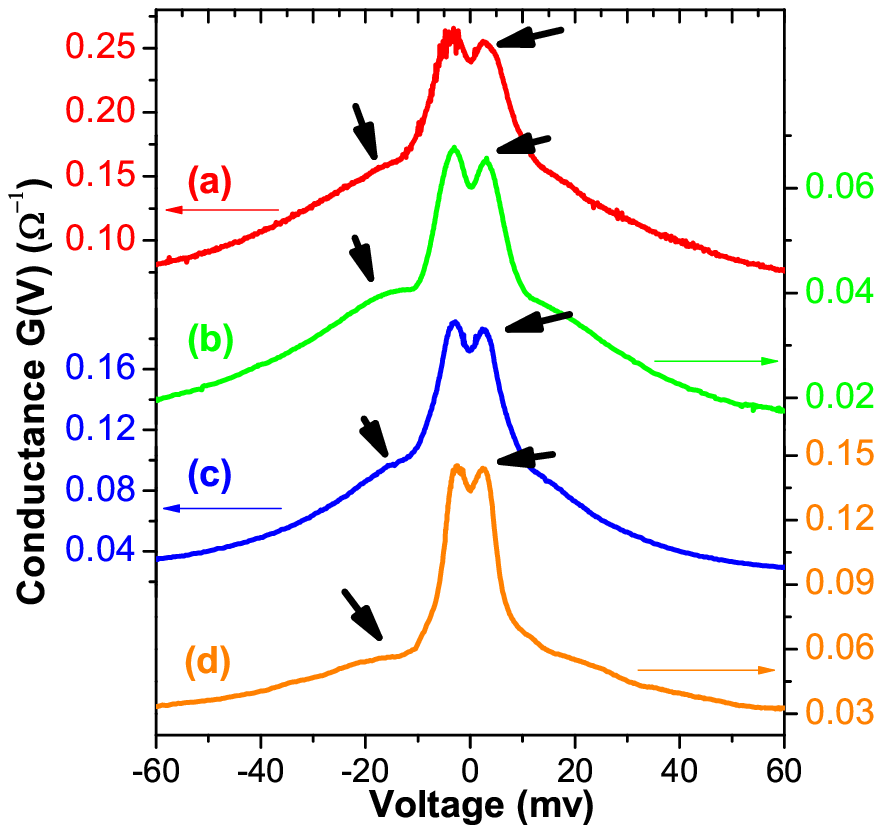}
 \caption{\label{Figure1} Differential conductance spectra G(V) for the
Au/(Ba$_{0.6}$K$_{0.4}$)Fe$_2$As$_2$ point-contact junctions at
low temperatures T$\sim$2 K. The peak and hump structures are
indicated by arrows nearby.}
 \end{figure}

\begin{figure}[!ht]
 \includegraphics[angle=0,width=0.95\textwidth]{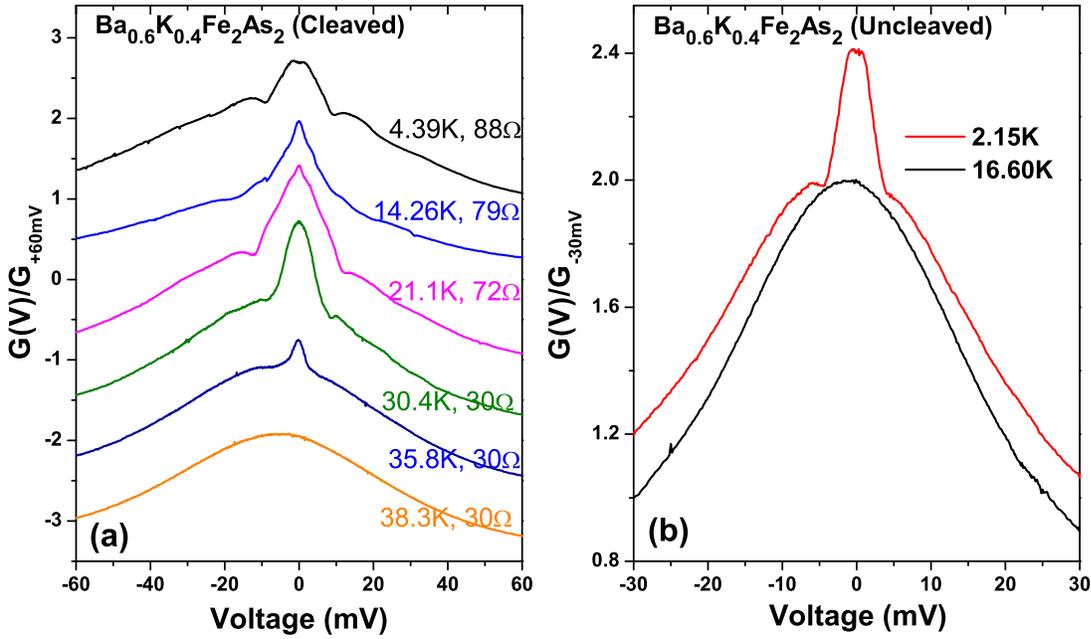}
 \caption{\label{Figure2a_2b} Temperature dependence of the
 conductance curves G(V) for the Au/(Ba,K)Fe$_2$As$_2$
point-contact junctions on a (a) fresh-cleaved surface (the
junction resistance changes with temperature due to instability of
the contact) and (b) uncleaved surface. The curves are vertically
shifted for clarity. }
 \end{figure}

For the (Ba,K)Fe$_2$As$_2$ crystals, representatives of the most
frequently observed G(V) curves at low temperatures ($\sim$2 K)
are shown in Figure \ref{Figure1}. The prominent features are
the two peaks at $\sim\pm3$ meV and a strongly sloping background.
A hump structure can also be noticed around $\pm 15$ meV as
indicated by arrows, and a small conductance asymmetry is
systematically observed. Similar broad backgrounds and
asymmetries are reported in the recent point-contact measurement
on SmFeAsO$_{0.8}$F$_{0.2}$, where the sloping background is
claimed to disappear around the Neel temperature ($\sim$ 140 K) of the
parent compound \cite{Gonnelli-3}. The temperature dependence as shown in Figure
\ref{Figure2a_2b}(a) verifies that the low-bias conductance
enhancement is due to Andreev reflection. Although the junction
resistance changes with temperature, the Andreev reflection
signal disappears only above the bulk $T_c$, giving confidence that
we probe the bulk gap. The sloping background survives above $T_c$
so it does not originate from superconductivity and must be due to
some other scattering mechanism.

Since the (Ba,K)Fe$_2$As$_2$ material is known to be reactive in air, we
 try to minimize exposure time between cleavage and cooldown. The
 usual time is about 30 minutes. We investigate the effect of a 1 week
 air exposure as shown in Figure \ref{Figure2a_2b}(b). The
 Andreev reflection signal is lost below 16.6 K, lower than
 that of the bulk. This indicates the air exposure degrades the
 surface and suppresses the superconductivity and that cleavage minimizes the surface degradation. We note the sloping
 background does not change with the air exposure.

\begin{figure}[!ht]
 \includegraphics[angle=0,width=0.95\textwidth]{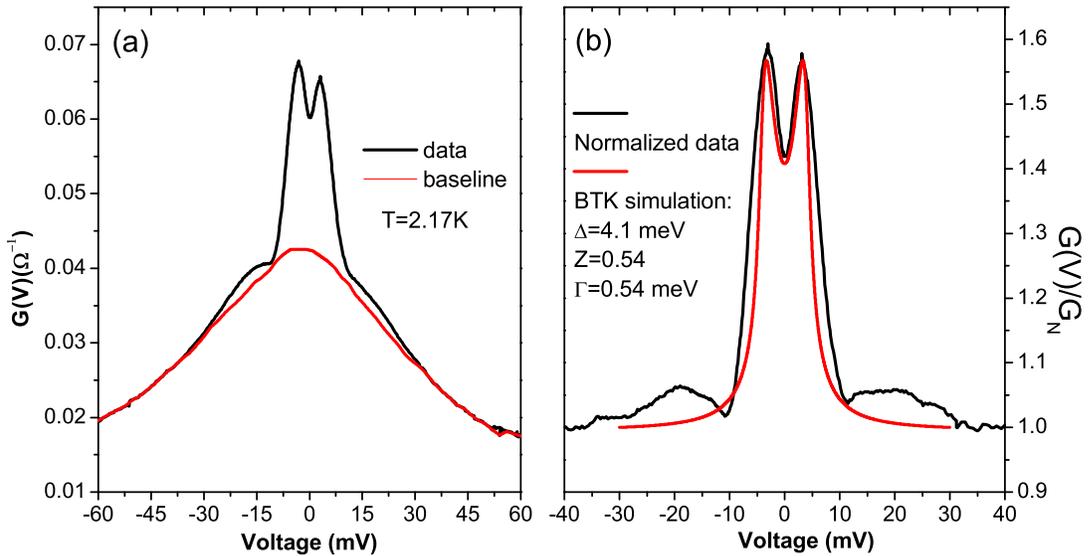}
  \caption{\label{Figure3a_3b}(a) Differential conductance curve G(V) of a Au/(Ba,K)Fe$_2$As$_2$
point-contact junction and extrapolated background baseline; (b)
The normalized conductance data and best fitting curve by BTK
model.}
 \end{figure}

For the gap analysis, we normalize the conductance for the cleaved surfaces to the
extrapolated baseline as shown in Figure \ref{Figure3a_3b}(a).
The normalized data are then analyzed by the single-gap
Blonder-Tinkham-Klapwijk (BTK) model \cite{blonder-1982-25,
Plecenik-1994-49}. The best fit for the energy gap is $\sim
3.0-4.0$ meV, so $2\Delta_0/k_BT_c \sim 2.0-2.6$, smaller than the
BCS weak coupling ratio of 3.52. This value is comparable with the
smaller gap size probed by PCARS on polycrystalline samples of LaFeAsO$_{1-x}$F$_x$
\cite{Gonnelli-1}, NdFeAsO$_{0.9}$F$_{0.1}$ \cite {PSamuely-1},
and SmFeAsO$_{0.8}$F$_{0.2}$ \cite {Gonnelli-3}. We stress that our materials are ``Fe-122" single crystals, and we probe in the c-axis orientation. This may account for us not observing the larger gap as follows: An ARPES study on (Ba,K)Fe$_2$As$_2$
reveals an isotropic but FS sheet-dependent gap structure, where
the $\beta$ band has an averaged gap size $\sim 5.8\pm0.8$ meV
with $2\Delta/k_BT_c\sim 3.6\pm0.5$ and the $\alpha$,$\gamma$ and
$\delta$ bands have comparable gap sizes around 11-13 meV with
$2\Delta/k_BT_c\sim 7.0-8.0$ \cite{HDing-2}. Considering the FSs
from band structure calculations, the $\alpha$,$\gamma$ and
$\delta$ bands are highly 2-dimensional with cylindrical shapes
while the $\beta$ band is strongly 3-dimensional due to the
$d_{3z^2-r^2}$ component \cite{GTWang-1}. The small gap observed
here may correspond to the 3D $\beta$ band. Because the Fermi
velocity on the $\alpha$,$\gamma$, and $\delta$ bands is mostly in the
ab plane and perpendicular to the c-axis, these bands contribute a
relatively small spectral weight for current flowing in the c-axis,
and the coherent peaks from these larger-gap bands are almost
absent in PCARS, similar to the case of MgB$_2$ \cite{Gonnelli-2}.
Both gaps should then be observable if junction current is flowing
within the ab plane. Szab\'{o} $\emph{et al.}$ report two
superconducting gaps for PCARS measurement in the ab plane with
$2\Delta_1/k_BT_c=2.5-4$ and $2\Delta_2/k_BT_c=9-10$
\cite{PSamuely-3}. However, Andreev reflection is totally absent
for their point-contact junctions in c-axis. This is distinct from
our results and may be due to the sample difference.

\begin{figure}[!ht]
 \includegraphics[angle=0,width=0.95\textwidth]{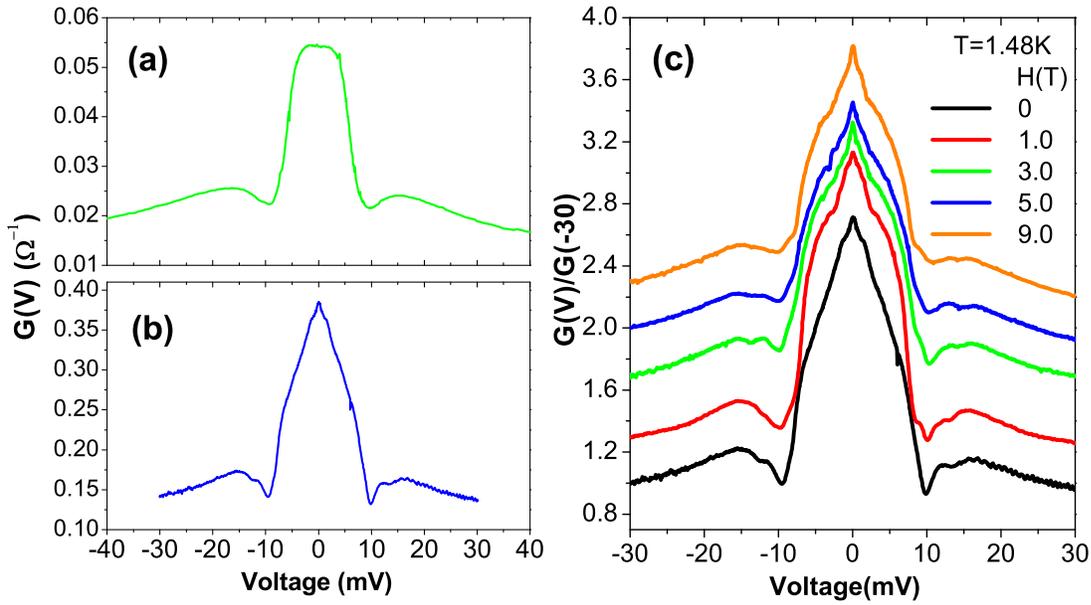}
 \caption{\label{Figure4a_4c} (a) $\&$ (b) Differential
conductance curves G(V) without double-peak structure for the
Au/(Ba,K)Fe$_2$As$_2$ point-contact junctions. (c) the field
dependence of the G(V) in (b) at T=1.48 K. The curves are
vertically shifted for clarity.}
 \end{figure}

The elastic and inelastic electron mean free paths, $l_{el}$ and $l_{in}$, respectively, are not known for (Ba,K)Fe$_2$As$_2$ single crystals. However, it is generally believed that
they are bad metals and \emph{l$_{el}$} could be a few tens of
nanometers, making it difficult to form a contact in the Sharvin
ballistic limit, (contact diameter d $<$ l$_{el}$).  For the
contact in the diffusive regime ($l_{el} < d <\sqrt{\emph{l$_{el}$}\emph{l$_{in}$}}$), dips may arise from the extra finite resistance of the superconducting electrode when
the junction current at a finite voltage bias exceeds the critical
current \cite{GSheet-1}. The humps in Figure \ref{Figure1} are a
possible signature of shallow dips, possibly dimmed by the sloping
background. As the point-contact is moved further away from the ballistic
regime, the dip structure becomes more pronounced
and a zero-bias conductance peak (ZBCP) is observed, rather
than the usual double-peak. This is likely the case as shown in
Figure \ref{Figure4a_4c}(a) \& (b) where a ZBCP is observed and the
dips at $\pm$ 10 meV are more pronounced than in Figure
\ref{Figure1}. Although a peak at zero bias may arise from Andreev
 bound states in a d-wave superconductor \cite{LHGreene-1}, the width, the lack of field
dependence as shown in Figure \ref{Figure4a_4c} (c), and the pronounced dip
structures together indicate the ZBCP is due to the contact not being in the
Sharvin limit.

\subsection{Ba(Fe,Co)$_2$As$_2$}
\begin{figure}[!ht]
 \includegraphics[angle=0,width=0.95\textwidth]{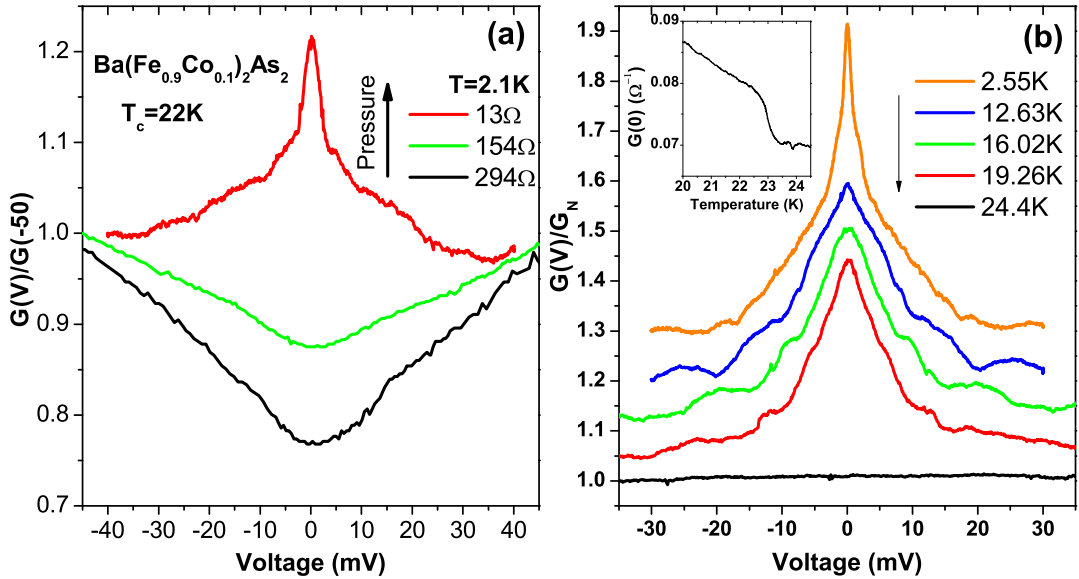}
 \caption{\label{Figure5a_5b} (a) The point-contact conductance curve,
  G(V), evolving from V-shape to ZBCP with the increase of tip pressure at T=2.1 K;
   (b) The temperature dependence of G(V) with ZBCP. The curves are vertically shifted
    for clarity. Inset shows the temperature evolution of the zero-bias conductance
    around the bulk T$_c$. }
 \end{figure}

 Conductance curves for the point-contact junctions on cleaved Ba(Fe,Co)$_2$As$_2$ single crystals are shown as a function of contact resistance and temperature in Figures \ref{Figure5a_5b}(a) and \ref{Figure5a_5b}(b), respectively. For most contacts, a conductance enhancement at zero-bias
with broad shoulders is observed. Note that with increasing temperature, the ZBCP
disappears at the bulk $T_c$ indicating it originates from Andreev
reflection. A V-shape G(V) is obtained when the tip is in gentle contact with the crystal hundreds of ohms junction resistance, and with increased tip pressure reducing the junction resistance, the conductance curve changes to the general ZBCP feature.

 \subsection{BaFe$_2$As$_2$, (Sr,Na)Fe$_2$As$_2$ \& Sr(Fe,Co)$_2$As$_2$}

\begin{figure}[!ht]
 \includegraphics[angle=0,width=0.95\textwidth]{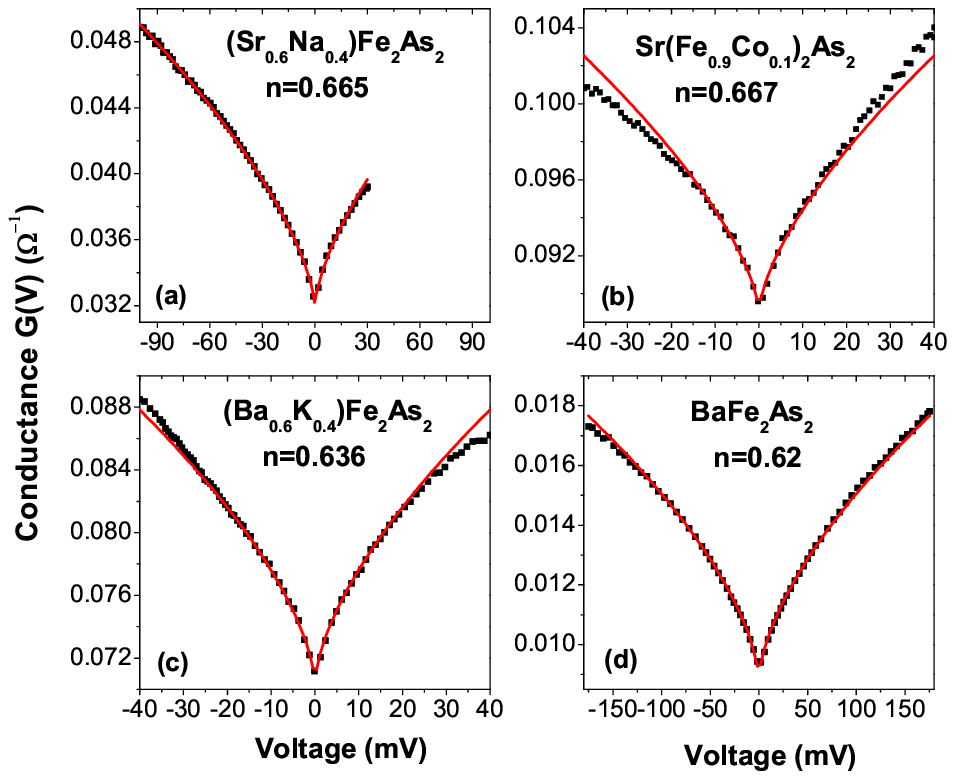}
 \caption{\label{Figure6a_6d} ZBA: G(V) curves observed (black) and power law fit (red)
 with n as the fitted power coefficient for the point-contact junctions on
(a)(Sr$_{0.6}$Na$_{0.4}$)Fe$_2$As$_2$; (b) Sr(Fe$_{0.9}$Co$_{0.1}$)$_2$As$_2$; (c)(Ba$_{0.6}$K$_{0.4}$)Fe$_2$As$_2$;
 (d) BaFe$_2$As$_2$. }
 \end{figure}

Figure \ref{Figure6a_6d} (a) and (b) show the typical G(V) curves with
reduced conductance at lower voltage bias observed for the point-contact
junctions on cleaved (Sr,Na)Fe$_2$As$_2$ and
Sr(Fe,Co)$_2$As$_2$ surfaces. Similar features are also reported by
other groups \cite{PSamuely-1,LCohen-1}. We find the conductance shape
does not change up to applied magnetic fields of 9 Tesla. There is no splitting of the
zero-bias anomaly (ZBA) in the magnetic field, which may rule out
the Kondo scattering as an origin. This ZBA feature, without
Andreev reflection, is probably due to the absence of
superconductivity in probed areas, even though the samples are
checked to be superconducting. Conductance curves with this ZBA, as shown in Figure
\ref{Figure6a_6d}(c), are sometimes observed in the point-contact
junctions on cleaved (Ba,K)Fe$_2$As$_2$ crystals. We have also
observed the same ZBA feature for the nonsuperconducting parent compound
BaFe$_2$As$_2$ single crystals even up to 200 meV, as shown in Figure
\ref{Figure6a_6d}(d), where SDW magnetic order exists. All the
curves can be fit to a power law function $G(V)=G(0)+c*|V|^n$
with a power coefficient, n $\sim 2/3$. This may indicate a
universal origin of this ZBA observed commonly among different
crystals.

In the BaFe$_2$As$_2$ parent compound, a structure
transition occurs at the antiferromagnetic ordering temperature, $T_N\sim$135 K. A phase-separated coexistence of magnetic order and superconductivity in Fe-122
family is reported in recent muon spin rotation ($\mu$SR) studies
\cite{JTPark-1, AAmato-1, TGoko-1} with a magnetic correlation length $>$100 \AA. Park $\emph{et al.}$ demonstrate the mesoscopic phase-separated coexistence of magnetically ordered and
non-magnetic states on a lateral scale of $\sim 65$ nm in the
slightly underdoped (Ba,K)Fe$_2$As$_2$ system \cite{JTPark-1}. Whether such coexistence is an intrinsic electronic property for Fe-122 system or due to some crystalline inhomogeneity remains uncertain.

In investigating the possibility that the ZBA originates from magnetic order, we note magnetic order can be detected by PCARS \cite{wkpark-3}. A point-contact junction made on a superconducting region would exhibit Andreev reflection and one on a nonsuperconducting, magnetically ordered region would not. Instead, a signature due to electron scattering from magnetic order may be detected. This may be the
case for (Sr,Na)Fe$_2$As$_2$ and Sr(Fe,Co)$_2$As$_2$ where no
Andreev reflection peak in G(V) has been observed, and G(V) curves with the ZBA are mostly observed. Goko et al. apply $\mu$SR to investigate (Ba,K)Fe$_2$As$_2$ and
(Sr,Na)Fe$_2$As$_2$ single crystals (same source as
ours) and argue that static magnetism sets in at temperatures well
above the superconducting T$_c$. They estimate the superconducting volume fraction to be 50 $\%$ in (Ba,K)Fe$_2$As$_2$ crystals and $\sim 90 \%$ in (Sr,Na)Fe$_2$As$_2$ crystals at low temperatures \cite{TGoko-1}. This is consistent with our more frequent observation of ZBA features in (Sr,Na)Fe$_2$As$_2$ than (Ba,K)Fe$_2$As$_2$ cystals.

Temperature dependent PCARS for the parent compound BaFe$_2$As$_2$ reveals the ZBA feature is broadened and reduced with increasing temperature. As the Neel temperature ($T_N\sim$ 135 K) is crossed, no dramatic change in the spectra is observed but thermal population effects may mask changes in conductance due to the magnetic transition.

\section{Summary}

In conclusion, our PCARS measurement on the cleaved
(Ba,K)Fe$_2$As$_2$ single crystal surfaces shows a superconducting
gap energy of $\sim$ 4 meV when G(V) curves are analyzed by the
single-gap BTK model. For point-contacts on Ba(Fe,Co)$_2$As$_2$,
zero-bias peak is frequently observed due to Andreev reflection. A universal power law behavior of the G(V)
observed for PCARS on different Fe-122 superconducting samples may
be a natural result of the mesoscopic phase-separated coexistence
of magnetic and superconducting phases in the 122 system. More
careful investigation is needed to understand the origin of the
power-law ZBA behavior. Point-contact measurement in the
ab-plane would be helpful to explore more
detailed gap structure of these new unconventional superconductors.  \\

\ack
 We thank R. Hasch for technical support. This work at UIUC is
supported by the U.S. DoE Award No. DEFG02-07ER46453 through the
Frederick Seitz Materials Research Laboratory and the Center for
Micro-analysis of Materials at UIUC. Xin Lu thanks the support
by NSF DMR 07-06013. HQY is support by the National Science Foundation of China (Grant No.10874146,10934005), the National Basic Research Program of China (Grant No.2009CB929104) and Zhejiang Natural Science Foundation (ZJNSF R0690113). The work at ORNL is supported by Division of
Materials Sciences and Engineering, Office of Basic Energy
Sciences, DOE.

\newpage

\newpage

\section*{References}
\begin{thebibliography}{00}

\bibitem {HHosono-1} Kamihara Y, Watanabe T, Hirano M and Hosono H 2008 {\it J. Am. Chem. Soc.} {\bf 130}  3296 
\bibitem {IMazin-1} Mazin I I, Singh D J, Johannes M D and Du M H 2008 {\it Phys. Rev. Lett.} {\bf 101} 057003  
\bibitem {HDing-1} Ding H \etal 2008 {\it Europhys. Lett.} {\bf 83} 47001  
\bibitem {HDing-2} Nakayama K \etal 2009 {\it Europhys. Lett.} {\bf 85} 67002 
\bibitem {AKaminski-1} Kondo T \etal 2008 {\it Phys. Rev. Lett.} {\bf 101} 147003 
\bibitem {XJZhou-1} Zhao L \etal 2008 {\it Chin. Phys. Lett.} {\bf 25} 4402 
\bibitem {Sebastian-1} Sebastian S E, Gillett J, Harrison N, Lau P H C, Mielke C H and Lonzarich G G 2008 {\it J. Phys.: Condens. Matter} {\bf 20} 422203  

\bibitem {KHashimot-1} Hashimoto K \etal 2009 {\it Phys. Rev. Lett.} {\bf 102} 017002 
\bibitem {AAAczel-1}  Aczel A A \etal 2008 {\it Phys. Rev.} B {\bf 78} 214503 
\bibitem {AJDrew-1} Drew A J \etal 2008 {\it Phys. Rev. Lett.} {\bf 101} 097010   
\bibitem {HHWen-3} Ren C, Wang Z S, Luo H Q, Yang H, Shan L and Wen H-H 2008 {\it Phys. Rev. Lett.} {\bf 101} 257006 
\bibitem {HHWen-1} Mu G, Luo H Q, Wang Z S, Shan L, Ren C and Wen H-H 2009 {\it Phys. Rev.} B {\bf 79} 174501 
\bibitem {RProzorov-1} Martin C \etal 2009 {\it Phys. Rev. Lett.} {\bf 102} 247002
\bibitem {RProzorov-2} Gordon R T \etal 2009 {\it Phys. Rev. Lett.} {\bf 102} 127004 
\bibitem {RProzorov-3} Gordon R T, Martin C, Kim H, Ni N, Tanata M A, Schmalian J, Mazin I I, Bud'ko S L, Canfield P C and Prozorov R 2009 {\it Phys. Rev.} B {\bf 79} 100506(R) 
\bibitem {NJCurro-1} Grafe H-J \etal 2008 {\it Phys. Rev. Lett.} {\bf 101} 047003 
\bibitem {KMatano-1} Matano K, Ren Z A, Dong X L, Sun L L, Zhao Z X and Zheng G-q 2008 {\it Europhys. Lett.} {\bf 83} 57001 
\bibitem{DParker-1} Parker D, Dolgov O V, Korshunov M M, Golubov A A and Mazin I I 2008 {\it Phys. Rev.} B {\bf 78} 134524 
\bibitem {HLuetkens-1} Luetkens H \etal 2008 {\it Phys. Rev. Lett.} {\bf 101} 097009  
\bibitem {Gonnelli-2} Gonnelli R S, Daghero D, Ummarino G A, Stepanov V A, Jun J, Kazakov S M and Karpinski J 2002 {\it Phys. Rev. Lett.} {\bf 89} 247004  
\bibitem {Gonnelli-Mgb2} Gonnelli R S, Daghero D, Calzolari A,
Ummarino G A, Dellarocca V and Stepanov V A 2004 {\it Phys. Rev.} B {\bf 69} 100504(R) 
\bibitem {Naidyuk-MgB2} Yanson I K and Naidyuk Yu G 2004 {\it Low Temp.
Phys.} {\bf 30} 261; 
\bibitem {wkpark-2} Park W K, Sarrao J L, Thompson J D and Greene L H 2008 {\it Phys. Rev. Lett.} {\bf 100} 177001 
\bibitem {CLChien-1} Chen T Y, Tesanovic Z, Liu R H, Chen X H and Chien C L 2008 {\it Nature} {\bf 453} 1224  
\bibitem {HHWen-2} Wang Y L, Shan L, Cheng P, Ren C and Wen H H 2009 {\it Supercond. Sci. Technol.} {\bf 22} 015018 
\bibitem {Gonnelli-1} Gonnelli R S, Daghero D, Tortello M, Ummarino G A, Stepanov V A, Kim J S and Kremer R S 2009 {\it Phys. Rev.} B {\bf 79} 184526 
\bibitem{PSamuely-1} Samuely P, Szab\'o P, Pribulov\'a Z, Tillman M E, Bud'ko S and Canfield P C 2009 {\it Supercond. Sci. Technol.} {\bf 22} 014003 
\bibitem{LCohen-1} Yates K A, Cohen L F, Ren Z-A, Yang J, Lu W, Dong X L and Zhao Z-X 2008 {\it Supercond. Sci. Technol.} {\bf 21} 092003 
\bibitem{Gonnelli-3} Daghero D, Tortello M, Gonnelli R S, Stepanov V A, Zhigadlo N D and Karpinski J 2008 arXiv: 0812.1141 

\bibitem{DMandrus-1} Sefat A S, Jin R, McGuire M A, Sales B C, Singh D and Mandrus D 2008 {\it Phys. Rev. Lett.} {\bf 101} 117004  

\bibitem{NLWang-1} Chen G L, Li Z, Dong J, Li G, Hu W Z, Zhang X D, Song X H, Zheng P, Wang N L and Luo J L 2008 {\it Phys. Rev.} B {\bf 78} 224512 

\bibitem{WKPark-1} Park W K and Greene L H 2006 {\it Rev. Sci. Instru.} {\bf 77} 023905  
\bibitem{XinLu-phd} Xin Lu, University of Illinois at Urbana Champaign, PhD thesis in preparation
\bibitem {blonder-1982-25} Blonder G E, Tinkham M and  Klapwijk T M 1982 {\it Phys. Rev.} B {\bf 25} 4515
\bibitem{Plecenik-1994-49} Plecen\'ik A, Grajcar M, Be\ifmmode \check{n}\else \v{n}\fi{}a\ifmmode \check{c}\else
\v{c}\fi{}ka  I C V, Seidel P and Pfuch A,  1994 {\it Phys. Rev.} B {\bf 49} 10016
\bibitem{GTWang-1} Wang G T, Qian Y, Xu G, Dai X and Fang Z 2009 arXiv:0903.1385  
\bibitem{PSamuely-3} Szab\'o P, Pribulov\'a Z, Prist\'a\u s G,
Bud'ko S L, Canfield P C and Samuely P 2009 {\it Phys. Rev.} B {\bf 79} 012503  
\bibitem{GSheet-1} Sheet G, Mukhopadhyay S and Raychaudhuri P 2004 {\it Phys. Rev.} B {\bf 69} 134507 
\bibitem{LHGreene-1} Aprili M, Badica E and Greene L H 1999 {\it Phys. Rev. Lett.} {\bf 83} 4630
\bibitem{JTPark-1} Park J T \etal 2009 {\it Phys. Rev. Lett.} {\bf 102} 117006 
\bibitem{AAmato-1} Amato A, Khasanov R, Luetkens H and Klauss H H 2009 arXiv: 0901.3139
\bibitem{wkpark-3} Park W K, Sarrao J L, Thompson J D, Pham L D, Fisk Z and Greene L H 2008 {\it Phys.} B {\bf 403} 731 
\bibitem{TGoko-1} Goko T \etal 2009 {\it Phys. Rev.} B {\bf 80}, 024508 

\end{thebibliography}
\end {document}